\documentclass[twocolumn,aps,pre]{revtex4-1}
\usepackage[T1]{fontenc}
\usepackage[latin9]{inputenc}
\usepackage{xcolor}
\usepackage{amsmath}
\usepackage{amssymb}
\usepackage{graphicx}
\usepackage{hyperref}
\PassOptionsToPackage{normalem}{ulem}
\usepackage{ulem}

\makeatletter

\providecolor{lyxadded}{rgb}{0,0,1}
\providecolor{lyxdeleted}{rgb}{1,0,0}

 \@ifundefined{textcolor}{}
 {%
   \definecolor{BLACK}{gray}{0}
   \definecolor{WHITE}{gray}{1}
   \definecolor{RED}{rgb}{1,0,0}
   \definecolor{GREEN}{rgb}{0,1,0}
   \definecolor{BLUE}{rgb}{0,0,1}
   \definecolor{CYAN}{cmyk}{1,0,0,0}
   \definecolor{MAGENTA}{cmyk}{0,1,0,0}
   \definecolor{YELLOW}{cmyk}{0,0,1,0}
 }

\usepackage{tabularx}
\usepackage{array}\usepackage{booktabs}\newcolumntype{C}{>{\centering\arraybackslash}X}

\makeatother

\begin{document}

\title{Quantum dot attached to superconducting leads: \\
Relation between symmetric and asymmetric coupling}

\author{A. Kadlecov\'a}

\affiliation{Department of Condensed Matter Physics, Faculty of Mathematics and
Physics, Charles University, Ke Karlovu 5, CZ-121 16 Praha
2, Czech Republic}

\author{M. \v{Z}onda}

\affiliation{Department of Condensed Matter Physics, Faculty of Mathematics and
Physics, Charles University, Ke Karlovu 5, CZ-121 16 Praha
2, Czech Republic}

\author{T. Novotn\'y}
\email{tno@karlov.mff.cuni.cz}
\affiliation{Department of Condensed Matter Physics, Faculty of Mathematics and
Physics, Charles University, Ke Karlovu 5, CZ-121 16 Praha
2, Czech Republic}

\date{\today}

\begin{abstract}
We study the Anderson single-level quantum dot attached to two BCS
superconducting leads with the same gap size. We reveal that a system
with asymmetric tunnel coupling to the leads ($\Gamma_{L}\neq\Gamma_{R}$)
can be related to the symmetric system with the same net coupling
strength $\Gamma=\Gamma_{L}+\Gamma_{R}$. Surprisingly, it is the
symmetric case which is the most general, meaning that all physical
quantities in case of asymmetric coupling are fully determined by
the symmetric ones. We give ready-to-use conversion formulas for the
$0-\pi$ phase transition boundary, on-dot quantities, and the Josephson
current, and illustrate them on the NRG results of Oguri, Tanaka
and Bauer {[}\emph{Phys. Rev. B} \textbf{87}, 075432 (2013){]} for the
three-terminal setup. We apply our theory to the recent $0-\pi$ transition
measurement of Delagrange et al. {[}\emph{Phys. Rev. B} \textbf{93},
196437 (2016){]} and determine the asymmetry of the experimental setup from the measured
transition width. Finally, we establish that the widely assumed Kondo
``universality'' of physical quantities depending only on the ratio
of the Kondo temperature and the superconducting gap $T_{K}/\Delta$
cannot hold for asymmetric junctions. 
\end{abstract}

\maketitle

\section{Introduction}
Superconducting quantum dot nanostructures generalizing
the conventional Josephson junctions have been subject of intensive
research in the past decade \cite{Wernsdorfer-2010,Rodero-2011}.
Versatility of the setup covering a wide range of gate-tunable nanostructures
\cite{Morpurgo-1999,Kasumov-1999,Kasumov-2003,Jarillo-2006,vanDam-2006,Jorgensen-2006,Cleuziou-2006,Jorgensen-2007,Grove-2007,Pallecchi-2008,Zhang-2008,Jorgensen-2009,Liu-2009,Eichler-2009,Winkelmann-2009,Pillet-2010,Katsaros-2010,Maurand-2012,Lee-2012,Pillet-2013,Kumar-2014,Delagrange-2015,Delagrange-2016, Xu-preprint}
promises great potential for applications but also allows for detailed
studies of their nontrivial physical properties including Josephson
supercurrent and Andreev subgap transport. In many cases the system
can be very well described by a simple single impurity Anderson model
(SIAM) coupled to BCS leads \cite{Luitz-2012} which, depending on
particular parameters, may exhibit so called $0-\pi$ transition signaled
by the sign reversal of the supercurrent \cite{vanDam-2006,Cleuziou-2006,Jorgensen-2007,Eichler-2009,Maurand-2012,Delagrange-2015,Delagrange-2016}
and crossing of Andreev bound states (ABS) at the Fermi energy \cite{Pillet-2010,Pillet-2013,Chang-2013}.
The $0-\pi$ transition is induced by the underlying impurity quantum
phase transition (QPT) related to the crossing of the lowest many-body
eigenstates of the system from a spin-singlet ground state with positive
supercurrent ($0$ phase) to a spin-doublet state with negative supercurrent
($\pi$ phase) \cite{Matsuura-1977,Glazman-1989,Rozhkov-1999,
Yoshioka-2000,Siano-2004,Choi-2004,
Sellier-2005,Novotny-2005,Karrasch-2008,Meng-2009,Luitz-2012}.

In this study, we address an aspect of the problem which has been
largely overlooked thus far, namely the systematic study of effects
of asymmetry of the coupling to the two superconducting leads. Asymmetric coupling is
quite generic in experiments, while theory typically addresses the
symmetric setup for simplicity (and lack of resources to cover many
asymmetric instances). In the normal nonequilibrium transport, the
symmetric setup is indeed just a special case in the continuum of
possibilities covering all asymmetries (while keeping the total coupling
constant). However, as we show here, for the superconducting SIAM
the symmetric case is the most general and all quantities for asymmetric
situations can be derived \emph{exactly} from it by simple analytical
prescriptions. Despite their simplicity, their nontrivial implications
concern both the fundamental properties of the model and the analysis
of experiments. 

\section{Symmetric-asymmetric relation}
\subsection{Theory} 
The single impurity Anderson model is described by the Hamiltonian 
\begin{subequations}
\begin{align}
\mathcal{H} & =\mathcal{H}_{{\rm dot}}+\sum_{\alpha}(\mathcal{H}_{{\rm lead}}^{\alpha}+\mathcal{H}_{T}^{\alpha}),
\end{align}
where $\alpha=L,R$ denotes the left and right superconducting leads
(and possibly additional normal-metal leads like
in Refs.~\cite{Governale-2008,*Futterer-2013,Oguri-2013,Paaske-2015,Domanski-2017}).
The first term 
\begin{equation}
\mathcal{H}_{{\rm dot}}=\epsilon\sum_{\sigma=\uparrow,\downarrow}d_{\sigma}^{\dagger}d_{\sigma}+Ud_{\uparrow}^{\dagger}d_{\uparrow}d_{\downarrow}^{\dagger}d_{\downarrow}
\end{equation}
describes an impurity with a single-particle level $\epsilon$. Operators
$d_{\sigma}^{\dagger},\,d_{\sigma}$ create (annihilate) an on-dot
electron with spin $\sigma$, and $U$ describes the local Coulomb
interaction. The BCS Hamiltonian of the superconducting leads reads 
\begin{equation}
\mathcal{H}_{{\rm lead}}^{\alpha}=\sum_{\mathbf{k}\sigma}\epsilon_{\alpha}(\mathbf{k})\,c_{\alpha\mathbf{k}\sigma}^{\dagger}c_{\alpha\mathbf{k}\sigma}-\Delta_{\alpha}\sum_{\mathbf{k}}(e^{i\varphi_{\alpha}}c_{\alpha\mathbf{k}\uparrow}^{\dagger}c_{\alpha\mathbf{\ -k}\downarrow}^{\dagger}+\textrm{H.c.}),
\end{equation}
where $c_{\alpha\mathbf{k}\sigma,\,}^{\dagger}c_{\alpha\mathbf{k}\sigma}$
are the creation and annihilation operators of electrons with momentum
$\mathbf{k}$ and spin $\sigma$, $\Delta_{\alpha}$ is the amplitude
of the superconducting gap, and $\varphi_{\alpha}$ is the superconducting
phase. Throughout this paper we will assume $\Delta_{L}=\Delta_{R}=\text{\ensuremath{\Delta}}$,
which is crucial for our derivation (and generic in experiments). 

The last term 
\begin{equation}
\mathcal{H}_{T}^{\alpha}=\sum_{\mathbf{k}\sigma}(t_{\alpha\mathbf{k}}c_{\alpha\mathbf{k}\sigma}^{\dagger}d_{\sigma}+\textrm{H.c.}),
\end{equation}
describes the coupling with $t_{\alpha}$ denoting the tunneling matrix
elements. We assume the tunnel-coupling magnitude $\Gamma_{\alpha}=\pi t_{\alpha}^{2}\rho_{\alpha}$
(where $\rho_{\alpha}$ is the normal-state density of lead electron
states at the Fermi level) to be constant in the energy range of interest.
We further denote $\Gamma\equiv\Gamma_{R}+\Gamma_{L}$.
\end{subequations}

The system can be described in the Matsubara formalism \cite{Mahan-2000,Zonda-2015,Zonda-2016}.
In the noninteracting $U=0$ case the Green function reads 
\begin{equation}
\widehat{G}_{0}(i\omega_{n})=\left(\begin{array}{cc}
i\omega_{n}\left[1+s(i\omega_{n})\right]-\epsilon & \Delta_{\varphi}(i\omega_{n})\\
\Delta_{\varphi}^{*}(i\omega_{n}) & i\omega_{n}[1+s(i\omega_{n})]+\epsilon
\end{array}\right)^{-1},
\end{equation}
where $\omega_{n}$ are fermionic Matsubara frequencies and the functions
$s(i\omega_{n})=\Gamma/\sqrt{\omega_{n}^{2}+\Delta^{2}}$ 
and $\Delta_{\varphi}(i\omega_{n})=(\Gamma_{L}e^{i\varphi_{L}}+\Gamma_{R}e^{i\varphi_{R}})\Delta/\sqrt{\omega_{n}^{2}+\Delta^{2}}$.

The interacting ($U\neq0$) system with selfenergy $\Sigma$ is then
described by the full Green function $\widehat{G}$, which is given
by the Dyson equation $\widehat{G}^{-1}(i\omega_{n})=\widehat{G}_{0}^{-1}(i\omega_{n})-\widehat{\Sigma}(i\omega_{n})$.
The Green function $\widehat{G}$ is known to be a functional of $\widehat{G}_{0}$,
only further depending on the interaction strength $U$\textcolor{magenta}{{}
}\cite{Mahan-2000}, meaning that model parameters $\Gamma_{\alpha}$,
$\varphi_{\alpha}$, $\epsilon$ only enter through $\widehat{G}_{0}$.
Any symmetries of the noninteracting system under the change of these
parameters are then preserved in the interacting case.\textcolor{magenta}{{}
}We will further deal with a correspondence of the case of symmetric
coupling ($\Gamma_{L}=\Gamma_{R}=\Gamma/2$) with an asymmetric case
($\Gamma_{L}\neq\Gamma_{R}$), accompanied
by a corresponding change in the $\varphi_{\alpha}$ parameters. 

We first notice that $\widehat{G}_{0}$ only depends on $\varphi_{L,R}$
through the off-diagonal part $\Delta_{\varphi}$ and, moreover, only through a frequency-independent factor $\Gamma_{L}e^{i\varphi_{L}}+\Gamma_{R}e^{i\varphi_{R}}$.
Introducing the superconducting phase difference $\varphi=\varphi_{L}-\varphi_{R}$,
$\varphi\in(-\pi,\pi)$, the average phase shift $\delta=(\varphi_{L}+\varphi_{R})/2$
and the coupling-parameter asymmetry $a\equiv\Gamma_{L}/\Gamma_{R}$,
we can now simplify the $\varphi$-dependent factor to 
\begin{subequations}
\begin{align}
  \Gamma_{L}e^{i\varphi_{L}}+\Gamma_{R}e^{i\varphi_{R}}= & \Gamma\sqrt{\mathrm{\chi}(\varphi,a)}e^{i(\delta+\Psi)},\label{eq:DeltaPhiAsym}\\
\intertext{with}  \mathrm{\mathrm{\chi}}(\varphi,a)= & 1-\frac{4a}{(a+1)^{2}}\sin^{2}\frac{\varphi}{2}\label{eq:fAsym}\\
\intertext{and the overall phase \ensuremath{\,\Psi}}  \Psi(\varphi,a)= & \arctan\left[\left(\frac{a-1}{a+1}\right)\tan\frac{\varphi}{2}\right].\label{eq:Faze_Psi}
\end{align}
\end{subequations}
\begin{figure*}
\includegraphics[width=0.95\textwidth]{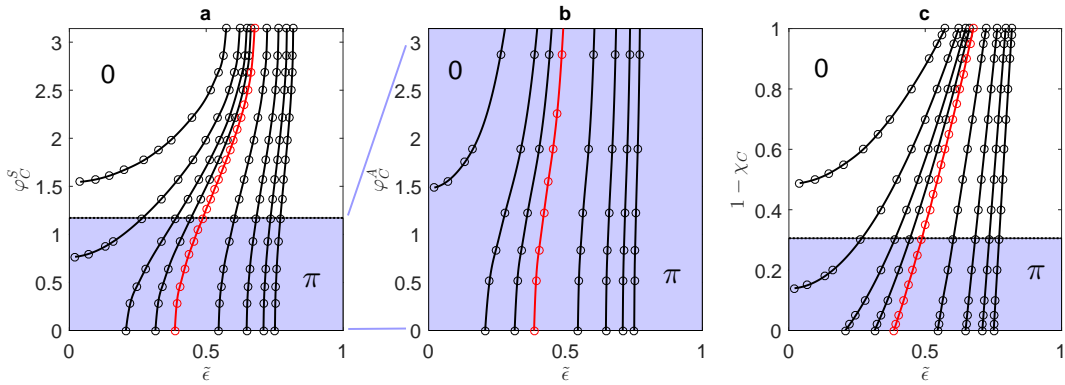} \caption{(Color online) $0-\pi$ phase boundary. (\textbf{\textcolor{black}{a}})
$\varphi_{C}^{S}$ dependence on the (shifted and normalized) energy
level $\tilde{\epsilon}\equiv1+2\epsilon/U$ in the symmetric case $\Gamma_{L}=\Gamma_{R}$.
The dots were obtained by NRG using the parameters $\Gamma=0.44$~meV, $\Delta=0.17$~meV and (from
left to right) $U=\{2,\,2.5,\,2.8,\,3,\,3.2,\,4,\,5,\,6,\,7\}$~meV. 
The red line ($U=3.2$~meV) represents the experimental values of Delagrange et al. \cite{Delagrange-2016}.
(\textbf{\textcolor{black}{b}}) $\varphi_{C}^{A}(\epsilon)$ \textemdash{}
plot \textbf{a} recomputed via Eq.~(\ref{eq:Phi_asym}) for asymmetry
$a=11$. The relevant $\varphi_{C}^{S}$ range is indicated by the
blue area in panels \textbf{\textcolor{black}{a}} and \textbf{\textcolor{black}{c}}. (\textbf{\textcolor{black}{c}}) $1-\chi(\varphi_{C})$ vs.~the shifted
and normalized dot level energy $\epsilon$. The curves are becoming
linear with increasing $U$.\label{fig:-phase-boundary}}
\end{figure*}

Equations \eqref{eq:DeltaPhiAsym}-\eqref{eq:Faze_Psi} make it
possible to relate an asymmetric junction to a symmetric one. 
Note that $\mathrm{\chi}(\varphi,a)$
is preserved by $a\rightarrow1/a$ (thus reflecting our freedom of
choice in the definition of the asymmetry) and reduces to $\chi(\varphi,1)\equiv\cos^{2}(\varphi/2)$
in the symmetric case. The value range of $\chi(\varphi,a=1)$ in the symmetric case is $\left[0,1\right]$, which shrinks to $[(a-1)^{2}/(a+1)^{2},1]$
for the asymmetric one ($a\neq1$). Together with continuity and monotony
of function $\chi(\varphi)$ this implies that for any physical (asymmetric) $\varphi^{A}$
there exist an effective (symmetric) $\varphi^{S}$ such, that 
\begin{equation}
\mathrm{\chi}(\varphi^{S},1)=\mathrm{\chi}(\varphi^{A},a).\label{eq:Podminkaprof}
\end{equation}
Inserting from Eq.~\eqref{eq:fAsym} we find
that 
\begin{subequations}
\begin{equation}
\begin{aligned}\varphi^{S} & =2\arccos\sqrt{\chi}=2\arccos\sqrt{1-\frac{4a}{(a+1)^{2}}\sin^{2}\frac{\varphi^{A}}{2}}.\end{aligned}
\label{eq:Phi_sym}
\end{equation}
To express $\varphi^{A}$ instead, we invert this formula and using \eqref{eq:fAsym} together with $2\arcsin x=\arccos(1-2x^{2})$ we obtain 
\begin{equation}
\begin{aligned}
\varphi^{A}&=\arccos\left(\frac{(a+1)^{2}}{2a}\left(\chi-1\right)+1\right)\\
&=\arccos\left(1 - \frac{(a+1)^{2}}{2a}\sin^2\frac{\varphi^S}{2}\right).
\end{aligned}
\label{eq:Phi_asym}
\end{equation}

Eqs.~\eqref{eq:Phi_sym} and \eqref{eq:Phi_asym} are substitution
relations for $\varphi$, which are sufficient to relate on-dot quantities.
To ensure correspondence of nonlocal quantities such as the Josephson
current, we also have to impose the condition 
\begin{equation}
\delta^{S}=\delta^{A}+\Psi(\varphi^{A}),\label{eq:DeltaS}
\end{equation}
\end{subequations}
which ties together the gauges of the symmetric and asymmetric cases.
Equations~\eqref{eq:Phi_sym} and \eqref{eq:DeltaS} fully describe the
correspondence, ensuring 
$\widehat{G}(\varphi^{A},\delta^{A})=\widehat{G}(\varphi^{S},\delta^{S})$.

We wish to stress, that counter-intuitively it is the \emph{symmetric}
case which contains all information and only needs to be examined
to fully understand the general (asymmetric) system. This is very
useful, because it allows us to calculate only characteristics of
the symmetric case and then compute all other cases from the formulas
above. Additionally, it is also a simple way to determine the asymmetry
of an experimental setup.

\subsection{Phase boundary}
We will now examine specific examples of the symmetric-asymmetric
relation, starting with the $0-\pi$ \emph{phase boundary}. In a suitable
range of parameters, when keeping $U$, $\Gamma$, and $\Delta$ fixed,
the state of the system depends on $\epsilon$ and
$\varphi$ and may exhibit the
$0-\pi$ transition. The transition curve is described by $\varphi_{C}(\epsilon)$,
$\varphi_{C}$ being the critical value at which the transition occurs
for a given $\epsilon$. Knowing the relation $\varphi_{C}^{S}(\epsilon)$ in the symmetric
case, we simply get the asymmetric transition curve by substitution in
Eq.~\eqref{eq:Phi_asym}. Figure~\ref{fig:-phase-boundary} shows
the phase diagram for different values of $U$, in symmetric and asymmetric
cases. The parameters $\Gamma=0.44$~meV, $\Delta=0.17$~meV correspond to
the experiment of Delagrange et al.~\cite{Delagrange-2016} 
 which will be addressed in more detail later on.
The symmetric curves in panel \textbf{\textcolor{black}{a}} were obtained
by numerical renormalization group (NRG) calculations \cite{Ljubljana-code,*ZitkoPruschke-2009}.
For small enough $U$ (2 and 2.5 meV in Fig.~\ref{fig:-phase-boundary}\textbf{a}) the $\pi$ phase exists only
above some finite critical $\varphi_C^S$ even at $\epsilon=-U/2$. 
The $\varphi_C^S$ curve changes to
approximately arccosine shape for larger $U$. Part \textbf{\textcolor{black}{b}} shows
the effect of asymmetry. The curves were obtained from panel \textbf{\textcolor{black}{a}}
by using the relation \eqref{eq:Phi_asym}. Only the lower part of
the symmetrical data in panel \textbf{\textcolor{black}{a}} is used,
and gets ``stretched'' by transformation \eqref{eq:Phi_asym} over
the whole $\varphi$ range (consequently, the phase boundary line
corresponding to $U=2$~meV disappears altogether). Panel \textbf{\textcolor{black}{c}}
shows a plot of $1-\chi(\varphi_{C})$ dependence on $\tilde{\epsilon}\equiv(\epsilon+U/2)/(U/2)=1+2\epsilon/U$. Due
to Eq.~\eqref{eq:Podminkaprof}, $\chi$ can be with advantage used
as an invariant variable.

\subsection{Physical quantities}
Physical quantities on the dot, like the free energy, particle
number, magnitude of the induced gap, or energy of the Andreev bound
states, which depend only on the local Green function in a gauge-invariant
way (i.e., are independent of $\delta$), can also be computed easily.
If the quantity's $\varphi$-dependence $F^{S}(\varphi^{S})$ is known
in the symmetric case, the symmetric-asymmetric relation reads $F^{A}(\varphi^{A})=F^{S}(\varphi^{S})$.
Inserting Eq.~\eqref{eq:Phi_sym} yields for any asymmetry $a$ the
equation 
\begin{equation}
\begin{aligned}F(\varphi) & =F^{S}\left(2\arccos\sqrt{1-\frac{4a}{(a+1)^{2}}\sin^{2}\frac{\varphi}{2}}\right).\end{aligned}
\label{eq:FodPhi}
\end{equation}
Here, $F(\varphi)$ and $\varphi$ are the physical quantities, that can be measured/tuned experimentally in a real asymmetric junction (we skip the superscript $A$ for physical quantities from now on). 

If we want to describe the \emph{Josephson current}, which is represented
by a nonlocal operator coupling the dot to lead(s), we encounter
a problem as its mean value is not determined from the Green's function
$\widehat{G}$ only but depends explicitly on $\varphi_{\alpha}$
as well \cite{Zonda-2016}. If one proceeds via this direct way, relation
\eqref{eq:DeltaS} must be also incorporated into the (rather tedious)
derivation as we explicitly demonstrate in Appendix \ref{appendixA}.
However, it is possible to follow an indirect route of the supercurrent
evaluation via the derivative $J\equiv 2e/\hbar\cdot \partial F/\partial\varphi$
of the free energy $F$, which satisfies the above symmetry relation
\eqref{eq:FodPhi}. Consequently, a prefactor appears in the symmetric-asymmetric
relation for the supercurrent: 
\begin{equation}
\begin{aligned}J(\varphi)=&\frac{\cos\frac{\varphi}{2}}{\sqrt{\frac{(a+1)^{2}}{4a}-\sin^{2}\frac{\varphi}{2}}}\times\\
& J^{S}\left(2\arccos\sqrt{1-\frac{4a}{(a+1)^{2}}\sin^{2}\frac{\varphi}{2}}\right).
\end{aligned}
\label{eq:Josephson_asym}
\end{equation}

\begin{figure}
\includegraphics[width=1\columnwidth]{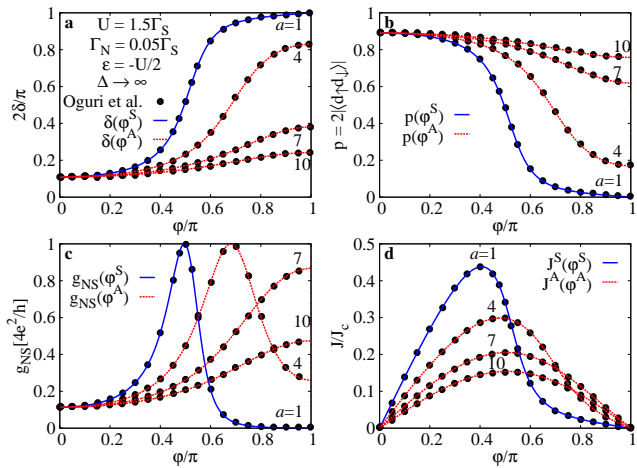}\caption{(Color online) Phase shift $\delta$ (\textbf{\textcolor{black}{a}}),
magnitude of the pair correlation (\textbf{\textcolor{black}{b}}),
Andreev conductance $g_{NS}$ (\textbf{\textcolor{black}{c}}) and
Josephson current $J$ in terms of $J_{C}=e\Gamma_{S}/\hbar$ (\textbf{\textcolor{black}{d}})
plotted as functions of the Josephson phase difference $\varphi$. The
black bullets have been taken graphically from Fig.~11 in Oguri \emph{et
al.~}\cite{Oguri-2013}. The blue solid lines have been calculated
for the symmetric coupling ($a=1$) using the NRG. The red dashed
lines, representing asymmetric coupling with $a=4,\,7,\,10$ have
been obtained from the blue ones using the relations
(\ref{eq:FodPhi}) (\textbf{\textcolor{black}{a}}-\textbf{\textcolor{black}{c}})
and (\ref{eq:Josephson_asym})~(\textbf{\textcolor{black}{d}}).\label{fig:Oguri}}
\end{figure}

Note that our theory can also be used if the setup contains one
or more normal metal lead(s) in addition to the two superconductors
like in Refs.~\cite{Governale-2008,*Futterer-2013,Oguri-2013,Paaske-2015,Domanski-2017}, because the normal metal leads do
not influence the physics of the superconducting phase difference.
This is illustrated in Fig.~\ref{fig:Oguri} containing different on-dot quantities
and the Josephson current for a setup with an additional normal electrode.
The black bullets were taken graphically from Fig.~11 of Ref.~\cite{Oguri-2013},
where they were calculated by NRG for symmetric as well as asymmetric
cases. On the other hand, we calculated by NRG only the blue lines
for the symmetric case, while red lines, corresponding to different
levels of asymmetry, were obtained by Eq.~\eqref{eq:FodPhi} for
on-dot quantities and Eq.~\eqref{eq:Josephson_asym} for the Josephson
current. With higher asymmetry, a smaller range of the symmetric data
is used. The on-dot quantities (a)-(c) are characteristically \textquotedbl{}stretched\textquotedbl{},
while the behavior of the Josephson current (d) is more complicated
due to the prefactor in Eq.~\eqref{eq:Josephson_asym}. In particular
$J(\varphi=\pi)=0$ also for the asymmetric cases. Our theory is in all
cases in excellent agreement with NRG results.

\subsection{Kondo (non)universality}

We wish to comment on the notion of Kondo universality
used, e.g., in Ref.~\cite{Delagrange-2016}. If the system is in the Kondo
regime  (big $U/\Gamma$ and small enough $\Delta/\Gamma$, see Fig.~\ref{fig:PhD}), physical quantities
are believed to only depend on $T_{K}/\Delta$, where
\begin{equation}
T_{K}=\sqrt{\frac{\Gamma U}{2}}\exp\left(-\frac{\pi\left|4\epsilon^{2}-U^{2}\right|}{8\Gamma U}\right)
\label{eq:Kondo}
\end{equation}
is the normal-state Kondo temperature. To our knowledge, in the superconducting
case universality has been tested both by numerical simulations of
a symmetric setting~\cite{Yoshioka-2000,Siano-2004,Choi-2004} and experimentally~\cite{Buizert-2007}.
As our example of the transition width demonstrates, physical quantities
can be altered by the asymmetry of the junction while keeping $\Gamma$
and, thus, also $T_{K}$ constant \textemdash{} recall the disappearance
of the transition line corresponding to $U=2$~meV ($U/\Gamma\approx 4.5$)
between panels \textbf{a} and \textbf{b} of Fig.~\ref{fig:-phase-boundary}.
Consequently, $T_{K}/\Delta$ scaling cannot hold for junctions with
different asymmetry and there is at least one more parameter to be
taken into account for any physical quantity $F$: $F^{S}(T_{K}/\Delta)\longrightarrow F(T_{K}/\Delta,a).$

\section{Analysis of the experiment}

\begin{figure}
\includegraphics[width=0.7\columnwidth]{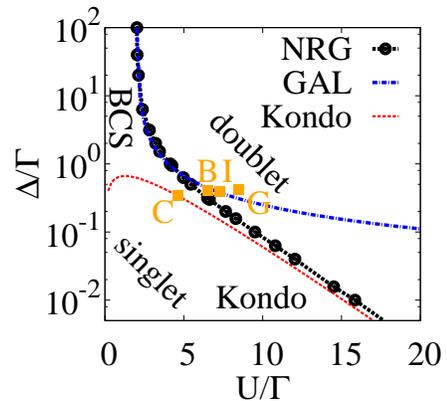}
\caption{Phase diagram in the $U-\Delta$ plane
of the superconducting single-impurity Anderson model with symmetric
leads at half-filling $\epsilon=-U/2$ and $\varphi=0$ (note the logarithmic scale on the vertical axis). Black line separates the singlet 
($0$ phase) and doublet ($\pi$ phase) ground states. 
The $0$ phase has two kinds of singlet ground state (BCS and Kondo) connected through a broad crossover region. NRG solution
(black bullets) is compared with the GAL approximation \cite{Zonda-2015,Zonda-2016} (blue dash-dotted line) and with an estimate of the critical delta from the Kondo temperature $\Delta_{C}\approx 1.37\, T_{K}$\cite{Yoshioka-2000} (red dashed line). The orange squares mark the positions of the
experimental setup taken from Ref.~\cite{Delagrange-2016}.
\label{fig:PhD}}
\end{figure}

One of the most beneficial outcomes of the symmetric-asymmetric relation is 
that all relevant experiments can be addressed with the symmetric models regardless of
the real coupling asymmetry. 
Moreover, Eq.~\eqref{eq:Phi_asym} can be used to obtain the value of the 
asymmetry from the experimental data as we now demonstrate.   
  
In their recent experimental study of a carbon nano tube (CNT) quantum dot,
Delagrange et al.~\cite{Delagrange-2016} focused on the $0-\pi$ transition controlled by the 
superconducting phase difference $\varphi$. Parameters of the single impurity Anderson model pertinent to the sample have been extracted from typical Coulomb diamonds appearing in the stability diagram and are summarized in Table~\ref{tab:asym}. The authors have successfully
fitted the $0-\pi$ phase transition curve $\varphi_{C}(\epsilon)$ on both sides of three Coulomb diamonds (called B, G, and I) with an arccosine dependence
\begin{equation}
\varphi_{C}=\arccos\left\{ -2\frac{\epsilon-\epsilon_{t}}{\delta\epsilon}\right\} \label{eq:PhiCAcosEpsilon}
\end{equation}
linear in energy. Here, $\delta\epsilon$ is the full width and $\epsilon_{t}$ is the position of the center of the transition curve.
This suggests an interesting universality, which, however, was not followed by the diamond C where  formula~(\ref{eq:PhiCAcosEpsilon}) had to be replaced by $\varphi_{C}\sim \arccos\left(c + \tilde{\epsilon}^2\right)$ \cite{Delagrange-2016}. This was reportedly because the transition took place close to $\tilde{\epsilon}=0$ and because it was ``incomplete", i.e., observed only above some finite critical $\varphi_{C}$ as in our Fig.~\ref{fig:-phase-boundary}. 

It should be mentioned that the arccosine functional form is necessary from the upper of Eq.~\eqref{eq:Phi_asym}; it is the {\em energy dependence} (linear or quadratic) which is nontrivial. To understand the energy-dependence behavior we first plot in Fig.~\ref{fig:PhD} a generic phase diagram in the $\Delta/\Gamma-U/\Gamma$ plane for $\varphi=0$, symmetric coupling to the leads, and at half-filling ($\epsilon=-U/2$; $\tilde{\epsilon}=0$). There are three regions: $\pi$ phase where the ground state is a spin doublet and $0$ phase with the BCS and Kondo singlet ground state regions \cite{Zonda-2015,Zonda-2016}. As can be seen the phase boundary calculated via NRG (black bullets) approaches in the Kondo region (small $\Delta$) the analytical curve \footnote{With $T_K$ from our Eq.~\eqref{eq:Kondo} which differs from $T_{K}$ used in Ref.~\cite{Yoshioka-2000} by a multiplicative factor $\approx 0.41$ and correspondingly changes the phase-boundary condition.} $\Delta_{C}\approx 1.37\, T_{K}$ (red line in Fig.~\ref{fig:PhD}), which was discussed in detail in Ref.~\cite{Yoshioka-2000}. On the other hand, we recently showed \cite{Zonda-2015,Zonda-2016} 
that the phase boundary in the BCS region and close to the half-filling can be very well approximated by a simple generalized atomic limit (GAL) formula 
$\chi_{C}=(U/2\Gamma)^2[1/(1+\Gamma/\Delta)^2-\tilde{\epsilon}^2]$ shown as the blue line in Fig.~\ref{fig:PhD} (with $\chi_{C}=1$ and $\tilde{\epsilon}=0$). We also depict as orange squares the experimental values of the  $\Delta$ and $U$ parameters normalized by total $\Gamma$ for diamonds B, C, G, and I discussed here (notice, that their placement in the phase diagram is only approximate as the phase transitions in those diamonds do not happen at the assumed half-filling).  

\begin{figure}
\includegraphics[width=1.0\columnwidth]{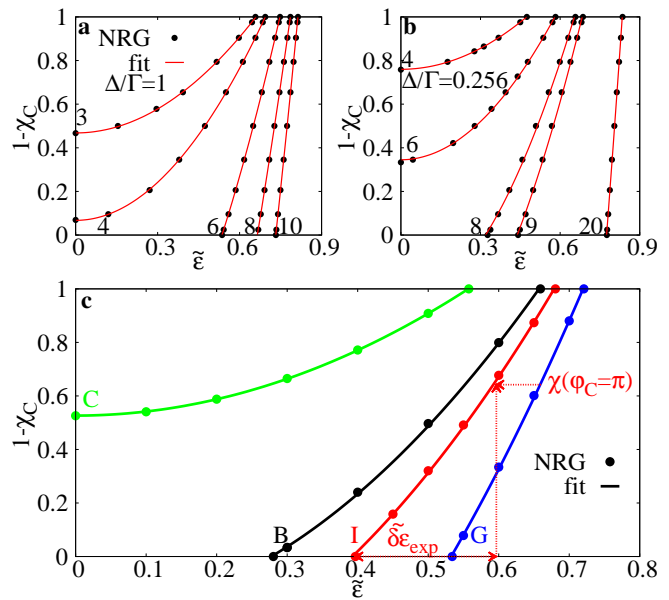}
\caption{Dependence of critical $\chi_{C}$ on $\tilde\epsilon$ calculated using NRG (dots) and fitted by 
the parabola Eq.~\eqref{eq:parabola}. ({\bf{a}}) For $\Delta/\Gamma=1$ the least-square fit error stayed below $0.2\%$ for all plotted values of $U/\Gamma$ (indicated by the numbers next to the curves).
({\bf{b}}) Case $\Delta/\Gamma=0.256$ had the biggest least-square fit error from all discussed cases; however, it was still below $1.2\%$ for all plotted values of $U/\Gamma$. ({\bf{c}}) Curves for all diamonds from the Table~\ref{tab:asym}.
As an example we marked the point $\chi\left(\varphi_C={\pi}\right)$ used in Eqs. (\ref{eq:a}) and (\ref{eq:chi_parabola})
for the right side of the diamond I. It's position was obtained using the measured value of $\widetilde{\delta\epsilon}_{\rm exp}$. 
\label{fig:parab_fit}}
\end{figure}

Interestingly, all the considered diamonds lie in the intermediate region where neither of the above two analytical formulas work (see the orange squares in Fig.~\ref{fig:PhD}). 
Nevertheless, we have discovered by exhaustive analysis of the NRG data that in compliance with GAL the critical $\chi_{C}(\tilde\epsilon)$ can be for sufficiently high $\Delta/\Gamma$ nearly perfectly fitted with a parabola
\begin{equation} 
\chi_{C}(\tilde\epsilon)=\gamma-\alpha \tilde{\epsilon}^2,
\label{eq:parabola}
\end{equation} 
although the coefficients $\alpha$ and $\gamma$ differ from the GAL values.
The examples of fits for different $\Delta/\Gamma$ and
$U/\Gamma$ are shown in Fig.~\ref{fig:parab_fit}.
We have found out that the least-square-fitting errors \footnote{We used a non-linear least-squares procedure within {\tt Gnuplot}, for the description of statistical errors see the documentation page \url{http://gnuplot.sourceforge.net/docs_4.2/node86.html}.} are below 
$0.5\%$ for both $\Delta/\Gamma=1$ (panel {\bf a}) as well as for $\Delta/\Gamma\sim0.3$
(panel {\bf c} corresponding to the experimental data from Fig.~\ref{fig:PhD})  
Even for the lowest considered ratio $\Delta/\Gamma=0.256$ (motivated by diamond A in Ref.~\cite{Delagrange-2016} not further discussed in here) was the 
difference from a parabola below $1.2\%$.
However, the parabolic dependence breaks down for still smaller values of  $\Delta/\Gamma$ (we have observed its clear breakdown already for $\Delta/\Gamma=0.1$; not shown) and it is definitely not valid in the strong Kondo regime. Yet, the above tests show that Eq.~(\ref{eq:parabola}) can be safely used for analysis of all considered diamonds.

This simple observation explains both above experimental findings. 
After inserting Eq.~(\ref{eq:parabola}) into Eq.~(\ref{eq:Phi_asym}),
which leads to 
\begin{equation}
\varphi_{C}=\arccos\left[\frac{(a+1)^{2}}{2a}\left(\gamma-\alpha \tilde\epsilon^2-\frac{a^2+1}{\left(a+1\right)^2}\right)\right],
\label{eq:fit_formula}
\end{equation}
one can immediately see that the seemingly anomalous transition for the diamond C 
with the quadratic energy dependence (see Fig.~\ref{fig:parab_fit}c) is actually quite generic.
On the other hand, the allegedly universal linear energy dependence of $\chi_{C}$ shown in Fig.~\ref{fig:-phase-boundary}c and
Fig.~\ref{fig:parab_fit} is in fact a limit of parabolas with large coefficients $\alpha$.

Parabolic character of the phase transition curve can be used to easily obtain the asymmetry of the experimental setups either by fitting the experimental data with Eq.~(\ref{eq:fit_formula}) or just by using a simple formula obtained from Eq.~(\ref{eq:fAsym}) (with the chosen solution $a>1$)
\begin{equation}
a=\frac{1+\sqrt{\chi\left(\varphi_C={\pi}\right)}}{1-\sqrt{\chi\left(\varphi_C={\pi}\right)}},
\label{eq:a}
\end{equation}
where $\chi(\varphi_C={\pi})$ can be read off from the theoretical curves using 
the normalized width of the transition $\widetilde{\delta\epsilon}_{\rm exp}\equiv2\delta\epsilon_{\rm exp}/U$ measured in the experiment as illustrated in Fig.~\ref{fig:parab_fit}c.
Alternatively, for complete transitions when $\varphi_C$ covers the whole $0$ to $\pi$ range (corresponding to large enough $U$ cases in Figs.~\ref{fig:-phase-boundary} and \ref{fig:parab_fit} ; B, G, and I diamonds in the experiment, see Fig.~\ref{fig:parab_fit}c) one 
can directly calculate $\chi_{C}(\varphi_{C}=\pi)$ from Eq.~\eqref{eq:parabola} as
\begin{equation}
\chi\left(\varphi_C={\pi}\right)=1-2\sqrt{\alpha(\gamma-1)}\widetilde{\delta\epsilon}_{\rm exp}-\alpha\widetilde{\delta\epsilon}_{\rm exp}^{2}.
\label{eq:chi_parabola}
\end{equation}
The big advantage of determining the asymmetry $a$ from formulas (\ref{eq:a}) and (\ref{eq:chi_parabola}) 
is that the transition width $\delta\epsilon$ is a robust quantity \cite{Zonda-2016,Delagrange-2016} and 
that coefficients $\alpha$ and $\gamma$ can be easily 
extracted from a few points (actually in the ideal case from just two)
of the phase boundary calculated for 
the symmetrical coupling using, e.g., NRG \footnote{One point can be chosen with $\varphi_{C}=0$ which is equivalent to a QD coupled to a {\em single} superconducting lead and amounts to a single-channel NRG problem that can be handled rather easily. The second point will necessarily involve a two-channel NRG problem which is more computationally demanding. Still the computational requirements boil down basically to a single set of two-channel NRG calculations to determine one phase-boundary point, which is very moderate.}.
     
In their experiment, Delagrange et al.~\cite{Delagrange-2016} have
determined the transition widths on both sides of the diamonds B, C, G, and I.
For the right side of diamond I the asymmetry
$a=4 $ was found via Quantum Monte Carlo (QMC) simulations \cite{Delagrange-2015};
for B, C, G, and the left side of I it remained unknown.  We have applied Eq.~\eqref{eq:a} to find the asymmetry of each diamond from the measured $\widetilde{\delta\epsilon}_{\rm exp}$ 
and coefficients $\gamma$ and $\alpha$ 
obtained by fitting the symmetric-coupling phase-boundaries $\chi_{C}(\epsilon)$ calculated via NRG shown in Fig.~\ref{fig:parab_fit}c (dots) with formula (\ref{eq:parabola}) (solid lines). The results are summarized in Table~\ref{tab:asym}. Analysis of the right side of diamond I (red lines in Figs.~\ref{fig:-phase-boundary} and~\ref{fig:parab_fit}c) with the measured transition width $\widetilde{\delta\epsilon}_{\rm exp}=0.20$ has given the asymmetry $a=4$, which agrees with the value obtained earlier via QMC \cite{Delagrange-2015}. The values of $a$ obtained for the diamond G and left sides of diamonds I and B point to even bigger asymmetries.
On the other hand the normalized transition width $\widetilde{\delta\epsilon}_{\rm exp}$ measured on the right side of diamond B suggests  a symmetric junction. The measured value ($\widetilde{\delta\epsilon}_{\rm exp}=0.43$) is actually even wider than the width calculated for the symmetric coupling ($\widetilde{\delta\epsilon}=0.39$), but the difference is within the  $10\%$ experimental uncertainty. For the diamonds B, I and G, where the phase transition curves are close to linear, we also present the results ($a_{\rm lin}$) of an alternative fitting procedure based on linearization, which was motivated by Eq.~\eqref{eq:PhiCAcosEpsilon}. This linearization procedure, which might be the only alternative in the strong Kondo regime, is discussed in detail in Appendix \ref{appendixB}. The asymmetries obtained by both methods are in good agreement.            

Unfortunately, the analysis of the remaining diamond $C$ is still problematic. Unlike in other diamonds, the phase transition for diamond $C$ is incomplete, i.e., it only exists above some finite critical value of $\varphi_{C}$ (Fig.~\ref{fig:parab_fit}c) as was also clearly observed in the experiment.
The measured values ($\widetilde{\delta\epsilon}_{\rm exp}$ = $0.87$ and $0.96$) are significantly bigger than the calculated width of the symmetric case ($\widetilde{\delta\epsilon}=0.6$). One can perhaps assume that the large difference could happen as a combination of the $10\%$ uncertainties in the estimation of the diamond's parameters and the error in the fitted transition width. In any case, the big value of $\widetilde{\delta\epsilon}_{\rm exp}$ hints at a symmetric or nearly symmetric junction.  
       
\begin{table}
\begin{tabular}{|c|cc|cc|cc|cc|cc|}
\hline
  &\multicolumn{2}{c|}{B} & \multicolumn{2}{c|}{C} & \multicolumn{2}{c|}{G} & \multicolumn{2}{c|}{I}\\

 &left&right&left&right&left&right&left&right\\
\hline
U/$\Gamma$  & \multicolumn{2}{c|}{6.5} & \multicolumn{2}{c|}{4.6} & \multicolumn{2}{c|}{8.5} & \multicolumn{2}{c|}{7.3}\\
\hline 
$\Delta/\Gamma$  & \multicolumn{2}{c|}{0.4} & \multicolumn{2}{c|}{0.34} & \multicolumn{2}{c|}{0.425} & \multicolumn{2}{c|}{0.39}\\
\hline
$\alpha$  & \multicolumn{2}{c|}{2.82} & \multicolumn{2}{c|}{1.53} & \multicolumn{2}{c|}{4.27} & \multicolumn{2}{c|}{3.26}\\
\hline
$\gamma$  & \multicolumn{2}{c|}{1.22} & \multicolumn{2}{c|}{0.47} & \multicolumn{2}{c|}{2.21} & \multicolumn{2}{c|}{1.50}\\
\hline
$\widetilde{\delta\epsilon}_{\rm exp}$ & 0.23& 0.43& 0.87& 0.96& 0.06& 0.06& 0.15& 0.2\\
\hline
$\boldsymbol{a}$ & \bf{5.7}& $\boldsymbol{1^{\#}}$ & $\boldsymbol{1^{\#}}$&$\boldsymbol{1^{\#}}$& \bf{11.8}& \bf{11.8}& \bf{6.6}& \bf{4.0}\\
\hline
$\beta_{\rm NRG}^A$ & 0.479 & \# & & & 0.202 & 0.202 & 0.347 & 0.317 \\
\hline
$a_{\rm lin}$ & 5.8& $1^{\#}$& & & 11.4& 11.4& 6.4& 4.0\\
\hline
\end{tabular}

\caption{Transition width and asymmetry \textemdash{} summary of experimental
data \cite{Delagrange-2016} and our corresponding results. Columns
correspond to the measured Coulomb diamonds. The rows give the measured normalized parameters $U/\Gamma$, $\Delta/\Gamma$, and coefficients
$\alpha$ and $\gamma$ of Eq.~\eqref{eq:parabola} obtained from the NRG data fitting.  Normalized transition widths $\widetilde{\delta\epsilon}_{\rm exp}\equiv2\delta\epsilon_{\rm exp}/U$ measured for both sides of the diamonds are followed by the asymmetry $a$ determined by the procedure discussed in the main text.
The last two lines are parameters (negative slope and asymmetry) obtained by an alternative linearization fitting procedure, which is discussed in Appendix \ref{appendixB}. (\#) Here, $\widetilde{\delta\epsilon}_{\rm exp}$ is bigger then the transition width of the symmetric case (see the main text for details).\label{tab:asym}}
\end{table}

\section{Conclusions}
We have unveiled the thus far unnoticed simple, yet
very powerful correspondence between the characteristics of a single-level
quantum dot coupled symmetrically or asymmetrically to two phase-biased
superconducting leads and potentially to further normal lead(s). We
have found that, counterintuitively, the symmetric setup is the most
general one and its knowledge allows full description of the equivalent
asymmetric system for any value of the asymmetry of the coupling.
This discovery enables to utilize known results for symmetric setups
in general asymmetric cases via trivial analytical relations. Moreover,
it also provides an efficient tool for estimating the coupling asymmetry
from the experimental data, which is otherwise a demanding task.

We have demonstrated the potential of this method by its application to recent experimental data in combination with a phenomenological analysis of the structure of phase-transition curves. We have discovered by exhaustive NRG calculations that the phase boundaries in a wide range of parameters (away from strong Kondo regime) are described by a simple quadratic functional dependence whose two parameters can be obtained with a moderate computational effort and utilized in a trivial analytical evaluation of the coupling asymmetry from the measured phase-transition width. In case of the single previously existing theoretical asymmetry prediction obtained by direct Quantum Monte Carlo simulations for the given experiment, our result is in perfect agreement at a fractional computational cost. 

\begin{acknowledgments}
This work is supported by the National Science Centre (Poland) through
the grant DEC-2014/13/B/ST3/04451, the Czech Science Foundation
via Project No.~16-19640S and the Charles University project GA UK No.~888217.
\end{acknowledgments}

\appendix
\section{Derivation of the symmetric-asymmetric relation for the Josephson current}\label{appendixA}
The main aim of Appendix \ref{appendixA} is to show a direct derivation of the symmetric-asymmetric relation for the Josephson current. Not only should this serve as an illustration of the procedure but also as a guide for the derivation of other quantities which depend not only on the Green function but also explicitly on $\varphi_\alpha$'s.

Starting with the \emph{interacting} Green function $\widehat{G}(i\omega_{n})$,
the Josephson current flowing into lead $\alpha$ can be expressed
as a sum over Matsubara frequencies~\cite{Zonda-2016} 
\begin{equation}
J_{\alpha}=4k_{B}T \sum_{\omega_{n}}\frac{\Gamma_{\alpha}\Delta}{\sqrt{\Delta^{2}+\omega_{n}^{2}}}\mathrm{Im}\left[\mathcal{G}(i\omega_{n})e^{-i\varphi_{\alpha}}\right]\label{eq:Current},
\end{equation}
where $T$ denotes the  temperature and $\mathcal{G}(i\omega_{n})$
is the off-diagonal element of $\widehat{G(}i\omega_{n})$. As this
expression explicitly depends on $\varphi_{\alpha}$, we will have
to handle the correspondence between the symmetric and asymmetric
case as having two parameters, the phase difference $\varphi=\varphi_{L}-\varphi_{R}$
and shift $\text{\ensuremath{\delta}=}(\varphi_{L}+\varphi_{R})/2$.
As shown in the main text, these are connected by the transformation
\begin{equation}
\begin{aligned}\varphi^{S} & =2\arccos\sqrt{\chi(\varphi^{A})},\\
\delta^{S} & =\delta^{A}+\Psi(\varphi^{A}),
\end{aligned}
\end{equation}
with $\mathrm{\mathrm{\chi}}(\varphi^{A})=1-\frac{4a}{(a+1)^{2}}\sin^{2}\frac{\varphi^{A}}{2}$,
and $\Psi(\varphi^{A})=\arctan\left[\left(\frac{a-1}{a+1}\right)\tan\frac{\varphi^{A}}{2}\right]$,
and asymmetry $a\equiv\Gamma_{L}/\Gamma_{R}$. We equivalently express
the parameters $\varphi^{S}_{L}$, $\varphi^{S}_{R}$ directly, obtaining
\begin{equation}
\begin{aligned}\varphi_{L}^{S} & =\frac{1}{2}(\varphi_{L}^{A}+\varphi_{R}^{A})+\Psi(\varphi^{A})+\arccos\sqrt{\chi(\varphi^{A})},\\
\varphi_{R}^{S} & =\frac{1}{2}(\varphi_{L}^{A}+\varphi_{R}^{A})+\Psi(\varphi^{A})-\arccos\sqrt{\chi(\varphi^{A})}.
\end{aligned}
\label{eq:phi_L_R}
\end{equation}
With this choice of symmetric-asymmetric relation the Green function
(and in particular it's off diagonal element) is preserved, so that
$\mathcal{G}(\varphi_{L}^{A},\varphi_{R}^{A})\equiv\mathcal{G}(\varphi_{L}^{S},\varphi_{R}^{S})$.
To compute the current we first denote 
\begin{equation}
\begin{aligned}\mathcal{I} & =2k_{B}T\sum_{\omega_{n}}\frac{\Gamma\Delta}{\sqrt{\Delta^{2}+\omega_{n}^{2}}}\mathrm{Im}\left[\mathcal{G}(i\omega_{n})\right],\\
\mathcal{R} & =2k_{B}T\sum_{\omega_{n}}\frac{\Gamma\Delta}{\sqrt{\Delta^{2}+\omega_{n}^{2}}}\mathrm{Re}\left[\mathcal{G}(i\omega_{n})\right].
\end{aligned}
\end{equation}
Then Eq.~\eqref{eq:Current} gives in the symmetric case 
\begin{equation}
J_{\alpha}^{S}=\mathcal{I\,}\cos\varphi_{\alpha}^{S}-\mathcal{R\,}\sin\varphi_{\alpha}^{S}.\label{eq:Jalpha}
\end{equation}
Here we have used $\Gamma=2\Gamma_{\alpha}$ and evaluated the imaginary
part of $\mathcal{G}(i\omega_{n})e^{-i\varphi_{\alpha}}$. From the
current conservation law $J=J_{L}=-J_{R}$ and Eq.~\eqref{eq:Jalpha}
we get two equations ($\alpha=L,R$) for $\mathcal{I}$ and $\mathcal{R}$
with $J$ as the parameter, leading to 
\[
\begin{aligned}\mathcal{I} & =-\frac{J^{S}}{\sin\varphi^{S}}\left(\sin\varphi_{R}^{S}+\sin\varphi_{L}^{S}\right),\\
\mathcal{R} & =-\frac{J^{S}}{\sin\varphi^{S}}\left(\cos\varphi_{R}^{S}+\cos\varphi_{L}^{S}\right).
\end{aligned}
\]

By using $\Gamma_{L}=\frac{a}{a+1}\Gamma$ in Eq.~\eqref{eq:Current},
the expression for the current incoming to the left lead for the asymmetric
case reads 
\begin{equation}
J^{A}=J_{L}^{A}=\frac{2a}{a+1}2k_{B}T\sum_{\omega_{n}}\frac{\Gamma\Delta}{\sqrt{\Delta^{2}+\omega_{n}^{2}}}\mathrm{Im}\left[\mathcal{G}(i\omega_{n})e^{-i\varphi_{L}^{A}}\right].\label{eq:J_L^A}
\end{equation}
Now we do the important step to express $J^{A}$ in terms of the corresponding
$J^{S}$. Since the quantities $\mathcal{R}$, $\mathcal{I}$ are
invariant under the transformation \eqref{eq:phi_L_R}, we can insert
them into \eqref{eq:J_L^A}. After simple rearrangement we obtain
\begin{equation}
J^{A}=-\frac{2a}{a+1}\frac{J^{S}(\varphi^{S})}{\sin\varphi^{s}}\left\{ \sin(\varphi_{L}^{S}-\varphi_{L}^{A})+\sin(\varphi_{R}^{S}-\varphi_{L}^{A})\right\} .
\end{equation}
Using the relations \eqref{eq:phi_L_R} to eliminate $\varphi_{L,R}^{S}$
yields the gauge invariant form
\begin{align}
\begin{aligned}J^{A}(\varphi^{A}) & =-J^{S}\frac{2a}{a+1}\,\frac{1}{\sin(2\arccos\sqrt{\chi})}\times\\
 & \left\{ \sin\left(\Psi-\frac{\varphi^{A}}{2}+\arccos\sqrt{\chi}\right)\right.\\
+ & \left.\sin\left(\Psi-\frac{\varphi^{A}}{2}-\arccos\sqrt{\chi}\right)\right\} .
\end{aligned}
\end{align}
This is simplified with use of the formula $\sin\alpha+\sin\beta=2\sin\frac{\alpha+\beta}{2}\cos\frac{\alpha-\beta}{2}$
in the numerator, and $\sin2\alpha=2\sin\alpha\cos\alpha$, $\sin\arccos\alpha=\sqrt{1-\alpha^{2}}$
in the denominator. 

\begin{align}
\begin{aligned}J^{A}(\varphi^{A})&=  -J^{S}\frac{2a}{a+1}\,\frac{\sin\left(\Psi-\frac{\varphi^{A}}{2}\right)}{\sin(\arccos\sqrt{\chi})}\\
&= -J^{S}\frac{2a}{a+1}\frac{\sin\Psi\cos\frac{\varphi^{A}}{2}-\sin\frac{\varphi^{A}}{2}\cos\Psi}{\sqrt{1-\chi}}.\label{eq:Jprubezne}
\end{aligned}
\end{align}
Next, to get rid of $\Psi$, we compute 
\begin{align}
\begin{aligned}\cos\Psi & =\frac{1}{\sqrt{1+\left(\frac{a-1}{a+1}\right)^{2}\tan^{2}\frac{\varphi^{A}}{2}}}=\frac{\cos\frac{\varphi^{A}}{2}}{\sqrt{\chi}},\\
\sin\Psi & =\frac{\left(\frac{a-1}{a+1}\right)\tan\frac{\varphi^{A}}{2}}{\sqrt{1+\left(\frac{a-1}{a+1}\right)^{2}\tan^{2}\frac{\varphi^{A}}{2}}}=\frac{\left(\frac{a-1}{a+1}\right)\sin\frac{\varphi^{A}}{2}}{\sqrt{\chi}}.
\end{aligned}
\end{align}
After inserting these relations, Eq.~\eqref{eq:Jprubezne} becomes
\begin{equation}
J_{L}^{A}=\frac{-\frac{2a}{a+1}\left(\frac{a-1}{a+1}-1\right)\sin\frac{\varphi^{A}}{2}\cos\frac{\varphi^{A}}{2}}{\sqrt{\chi}\sqrt{1-\chi}}J^{S}.
\end{equation}
Since the numerator yields $(\chi-1)\cot\frac{\varphi^{A}}{2}$ (for
$\varphi^{A}\neq0$), the relation between the Josephson current in
the symmetric and asymmetric case is simplified to 
\begin{equation}
J^{A}(\varphi^{A})=\sqrt{\frac{(1-\chi)}{\chi}}\mathrm{\cot}\frac{\varphi^{A}}{2}J^{S}(2\arccos\sqrt{\chi}),
\end{equation}
or, explicitly in $\varphi$ 
\begin{equation}
\begin{aligned}J^{A}(\varphi) & =\frac{\cos\frac{\varphi}{2}}{\sqrt{\frac{(a+1)^{2}}{4a}-\sin^{2}\frac{\varphi}{2}}}\times\\
\times & J^{S}\left(2\arccos\sqrt{1-\frac{4a}{(a+1)^{2}}\sin^{2}\frac{\varphi}{2}}\right),
\end{aligned}
\end{equation}
which is Eq.~\eqref{eq:Josephson_asym}.

\section{Linearization and alternative fitting procedure}\label{appendixB}
 
\begin{figure}
\includegraphics[width=1\columnwidth]{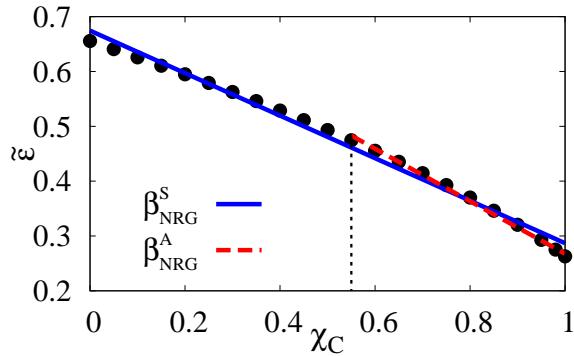}
\caption{(Color online) Illustration of the fitting procedure used to obtain the asymmetry.   
The numerically determined $\tilde{\epsilon}(\chi_{C})$ dependence (bullets) is approximated by a linear fit in the symmetric (blue solid line) and asymmetric (red dashed line) case.
The negative slopes are denoted $\beta^S_{\rm NRG}$ and $\beta^A_{\rm NRG}$ respectively. The fitting range for the asymmetric case is marked by the vertical dotted line.
In case of perfect linearity the lines would merge, but for a realistic curve on the verge of the Kondo regime they differ slightly.
\label{fig:fitting-scheme}}
\end{figure}

In this Appendix \ref{appendixB} we present an alternative analysis of the experiment based on linear approximation, which was motivated by the successful use of Eq.~\eqref{eq:PhiCAcosEpsilon} by Delagrange et al.~\cite{Delagrange-2016} in most of the measured cases (diamonds B, G, and I). We show that, even though the parabolic fit discussed in the main text is more general, the simpler linear approximation may give good enough results. Moreover, the linear approximation might be relevant in the fully developed Kondo regime.

Applying Eq.~(\ref{eq:Phi_asym}) to the phase boundary clearly shows that
$\varphi_{C}(\epsilon)$ has an exactly arccosine shape assumed in Eq.~\eqref{eq:PhiCAcosEpsilon}
if and only if $\chi_{C}(\epsilon)$ is linear in $\epsilon$. Indeed, from NRG data we observe that for big enough $U/\Gamma$ the phase boundary approaches a straight line, see Figs.~\ref{fig:-phase-boundary}c and \ref{fig:parab_fit}. Moreover, when asymmetry
is involved, only the values of $\chi_{C}$ close to $1$ (blue area in Fig.~\ref{fig:-phase-boundary}c) are used. This means that for large asymmetry the linear approximation becomes relevant for a wider range of parameters. 

We therefore assume $\mathrm{\chi_{C}(\tilde\epsilon)}\approx \kappa-\tilde\epsilon/\beta$. Inserting it into Eq.~(\ref{eq:Phi_asym}) gives (cf.~Eq.~\eqref{eq:fit_formula})
\begin{equation}
\varphi_{C}=\arccos\left[-\frac{(a+1)^{2}}{2a}\left(\frac{\tilde\epsilon}{\beta}+\frac{a^2+1}{\left(a+1\right)^2}-\kappa\right)\right].
\label{eq:PhiCAsym_lin}
\end{equation}
Comparison with Eq.~(\ref{eq:PhiCAcosEpsilon}) leads to 
\begin{align}
\widetilde{\delta\epsilon} & =\frac{4a}{(a+1)^{2}}\beta(\epsilon, U, \Gamma, \Delta),\label{eq:DeltaEpsilonLin}
\end{align}
where we have explicitly stated the dependence of the slope $\beta$ on given model parameters. 
For the ideal case of a perfectly linear dependence (or large enough $U$, see Fig.~\ref{fig:-phase-boundary}c), 
Eq.~\eqref{eq:DeltaEpsilonLin} separates the asymmetry dependence of the transition width from a universal (i.e., asymmetry-independent) slope $\beta$ (which also equals the transition width in the symmetric case). For experimentally relevant intermediate $U$'s (red curve in Fig.~\ref{fig:-phase-boundary}) 
the curve slightly bends and linear regression restricted to the appropriate range of $\chi_{C}$ is more precise and is used in the next paragraphs to analyze the experimental data.

We have applied Eq.~\eqref{eq:DeltaEpsilonLin} to find the asymmetry of diamonds B, G, and I from the measured transition width $\widetilde{\delta\epsilon}_{\rm exp}$ and the theoretical slope $\beta_{\rm NRG}$ obtained by performing a linear fit of phase-boundary curves $\chi_{C}(\tilde\epsilon)$ calculated via NRG. To account for their weak nonlinearities, we restrict the fitting range of $\text{\ensuremath{\chi_C}}$ only to values $[(a-1)^{2}/(a+1)^{2},1]$ relevant for the asymmetric case (blue area in Fig.~\ref{fig:-phase-boundary}). The range (influencing the slope $\beta_{\textrm{NRG}}$) and the asymmetry $a$ were determined self-consistently as values matching most closely the experimentally measured transition widths. Figure~\ref{fig:fitting-scheme} features 
the $\tilde{\epsilon}(\chi_{C})$ dependence (bullets) calculated by NRG for the diamond B 
in the experiment of Delagrange et al.~\cite{Delagrange-2016}.
We have chosen the diamond B as an illustration, because in this case the linear approximation
used to determine the asymmetry is the least accurate (see Fig~\ref{fig:parab_fit}c).   

We performed two linear fits. The negative slope denoted as $\beta_{\textrm{NRG}}^{A}$ (red dashed line) was obtained 
using the restricted interval (marked by the vertical dashed line) as discussed in the main text.  
For comparison, we also include the fit in the whole range $[0,1]$ of $\chi_{C}$  with the negative slope denoted $\beta_{\textrm{NRG}}^{S}$ (blue solid line). For a realistic phase-boundary curve the two slopes are close but not completely identical.
The values of the $\beta_{\textrm{NRG}}^{A}$ used to determine the asymmetry are tabulated in Table \ref{tab:asym} in the main text, 
while corresponding values of $\beta_{\textrm{NRG}}^{S}$  for diamonds B, G, and I respectively are 0.388, 0.190, and 0.291. These values are equal to the transition width of a symmetric ($a=1$) junction. The estimated standard error of the fit of all slopes was less than $2\%$. Results for the asymmetry are also included in Table \ref{tab:asym} in the main text, and are in good agreement with values obtained by the quadratic approximation. 

\bibliographystyle{apsrev4-1}
\bibliography{Josephson}

\end{document}